\documentclass[aps,prc,twocolumn,showpacs,floatfix,superscriptaddress,nofootnoteinbib,longbibliography]{revtex4-1}

\usepackage{amsmath}
\usepackage{inputenc}
\usepackage{amssymb}
\usepackage{graphicx}
\usepackage[version=4]{mhchem}
\usepackage{braket}
\usepackage{amsfonts}
\usepackage{amssymb}
\usepackage{cancel}
\usepackage{mathtools}
\usepackage{bbold}
\usepackage{xcolor}
\usepackage{epsfig}
\usepackage{bm}
\usepackage{color}
\usepackage{float}
\usepackage{dcolumn}
\usepackage{multirow}
\usepackage{siunitx}
\usepackage[unicode=true,
  linktocpage,
  linkbordercolor={0.5 0.5 1},
  citebordercolor={0.5 1 0.5},
  colorlinks=true,
  linkcolor=blue,
  citecolor=blue,
  urlcolor=blue]{hyperref}  
\usepackage{cgp_macros}
\usepackage{lipsum} 
\usepackage{placeins}  

\graphicspath{{.}}

\newcommand{\beq}{\begin{equation}}
\newcommand{\eeq}{\end{equation}}
\newcommand{\beqa}{\begin{eqnarray}}
\newcommand{\eeqa}{\end{eqnarray}}

\newcommand{\NNLOsat}{NNLO$_{\rm sat}$}
\newcommand{\NNLOgod}{$\Delta$NNLO$_{\rm GO}$(450)}
\newcommand{\NNLOgodlow}{$\Delta$NNLO$_{\rm GO}$(394)}

\begin{document}

\title{Ab-initio coupled-cluster calculations of ground and dipole excited states in \ce{^8He}} %

\author{F.~Bonaiti}
\affiliation{Institut f\"ur Kernphysik and PRISMA$^+$ Cluster of Excellence, Johannes Gutenberg-Universit\"at, 55128
  Mainz, Germany}

\author{S.~Bacca}
\affiliation{Institut f\"ur Kernphysik and PRISMA$^+$ Cluster of Excellence, Johannes Gutenberg-Universit\"at, 55128
  Mainz, Germany}
\affiliation{Helmholtz-Institut Mainz, Johannes Gutenberg-Universit\"at Mainz, D-55099 Mainz, Germany}

\author{G.~Hagen}

\affiliation{Physics Division, Oak Ridge National Laboratory,
Oak Ridge, TN 37831, USA}
\affiliation{Department of Physics and Astronomy, University of Tennessee,
Knoxville, TN 37996, USA}

\begin{abstract}
\noindent
 We perform coupled-cluster calculations of ground- and dipole excited-state properties of the  \ce{^8He} halo nucleus with nucleon-nucleon and three-nucleon interactions from chiral effective field theory, both with and without explicit delta degrees of freedom. By increasing the precision in our coupled-cluster calculations via the inclusion of leading order three-particle three-hole excitations in the cluster operator, we obtain a ground-state energy and a charge radius that are consistent with experiment, albeit with a slight under-binding. We also investigate the excited states induced by the  electric dipole operator and present a discussion on the Thomas-Reiche-Kuhn and cluster sum rules. Finally, we compute the electric dipole polarizability, providing a theoretical benchmark for future experimental determinations that will study this exotic nucleus.
\end{abstract}

\maketitle

\section{Introduction}
\noindent
Light nuclei close to  the driplines exhibit fascinating phenomena, such as the formation of diluted structures where a tightly-bound core is surrounded by a halo of one or more weakly-bound nucleons. Signatures for halo structures are both a small separation energy and a large matter radius, that does not follow the typical $A^{1/3}$ behaviour characterising stable nuclei.
Since their discovery in the 1980s~\cite{tanihata1985,TANIHATA1988592}, halo nuclei have attracted a lot of attention in the nuclear physics community, both from the experimental and the theoretical point of view. While from the experimental side enormous progress has been made and
precision measurements of masses and radii are nowadays possible even for very short lived systems (see e.g., Ref.~\cite{Blaum_2013}), halo nuclei still represent an arduous task for nuclear theory. The reason lies in  their extended size, which challenges several of the available many-body methods.

The  Helium isotope chain is particularly interesting in terms of the physics of halo nuclei. The chain presents a unbound-bound staggering when adding an odd-even number of neutrons on top of $^4$He:
$^5$He is  unbound,  $^6$He is a bound halo nucleus, $^7$He is unbound, and  $^8$He is again a bound  halo nucleus. Of the two halo nuclei, $^6$He is a borromean  halo system~\cite{ZHUKOV1993151}, and $^8$He
is the only known four-neutron halo nucleus.
Both $^6$He and $^8$He  have already been extensively investigated in the literature also with the so called ab initio methods, see, e.g.,~\cite{hagen2007d,bacca2009,bacca2012,maris2012,caprio2014,Carolina,launey2020,holl2021}.

In this paper, we will focus on the $^8$He halo nucleus and present a study based on the ab-initio coupled-cluster (CC) method~\cite{coester1958,coester1960,cizek1966,cizek1969,kuemmel1978,bartlett2007,hagen2013c}. Two reasons motivate this analysis. On the one hand, $^8$He can be seen as the most exotic  nucleus having the largest neutron-to-proton ratio across the nuclear chart ($N/Z = 3$), and as such it is interesting to test the models of nuclear forces developed in the ab initio community on this nucleus. On the other hand, there have been some recent
updates on the experimental determinations of the $^8$He ground state properties, for example of its charge radius~\cite{krauth2021}, and  measurements of its excitation spectrum
have either been made~\cite{holl2021} or are being planned~\cite{aumann}. Hence, new calculations based on the most modern interactions and many-body methods are interesting.


Our starting point to describe the nucleus of $^8$He
is the intrinsic nuclear Hamiltonian
\begin{equation}
\label{H}
H = \frac{1}{2mA} \sum_{i<j}^A (\vec{p}_i -\vec{p}_j)^2 + \sum _{i<j}^A V_{ij} + \sum_{i<j<k}^A W_{ijk}  \,,
\end{equation}
where $m$ is the nucleon mass, $A=8$ is the mass number,
$V_{ij}$ is the two-body force and $W_{ijk}$ is the three-body force. In the last years, a lot of progress has been achieved in deriving two- and three-body forces from chiral effective field theory ($\chi$EFT)~\cite{epelbaum2009,machleidt2011,hammer2019}, and different optimization strategies have been implemented for the low-energy constants (LECs)~\cite{ekstrom2013,ekstrom2015a}. In particular, chiral interactions with explicit delta degrees of freedom are also becoming available~\cite{kaiser1998,krebs2007,epelbaum2008,piarulli2015,ekstrom2017,jiang2020}. In this work we will explore both delta-full and delta-less interactions in our computations of $^8$He.

We will focus on ground state properties such as the binding energy and the charge radius, and on the low-energy excited states of dipole nature. For the latter, we will perform calculations based on a method that merges the Lorentz integral transform (LIT) with CC theory, called LIT-CC~\cite{bacca2013,bacca2014}. This approach has already proved to be successful in capturing  properties of unstable neutron-rich nuclei, such as \ce{^{22}O} in Ref.~\cite{bacca2014}, where the pygmy dipole resonance  was reproduced  using a chiral two-body interaction. With respect to Ref.~\cite{bacca2014}, we now have the advantage that we are able to include three-nucleon forces and effects of triples correlations in the CC expansion~\cite{miorelli2018}.

The paper is organized as follows. In Section \ref{computational_tools}, a description of our theoretical approach is provided. In Section \ref{results}, we present an overview of our results, separating the ground-state observables from the dipole excited states properties. For the latter  we devote one subsection to the energy-weighted sum rule and one subsection to the electric dipole polarizability. In Section \ref{conclusions}, we draw our conclusions.

\section{Computational tools}
\label{computational_tools}
\noindent
For a given nuclear Hamiltonian H, the CC approach is based on an exponential ansatz for the nuclear many-body wavefunction
\begin{equation}
    \ket{\Psi_0} = e^{T} \ket{\Phi_0}\,.
\end{equation}
Here, $\ket{\Phi_0}$ is a reference Slater determinant state, typically obtained from a Hartree-Fock calculation, where  single-particle states are projected onto the harmonic oscillator (HO) basis. The cluster operator $T$ introduces correlations in the many-body wavefunction, and  can be expanded in terms of $n$-particle $n$-hole excitations as
\begin{equation}
T = T_1 + T_2 + T_3 + \dots + T_A.
\end{equation}
The Schr\"odinger equation for the ground state of an $A$-particle system can be rewritten in the following form
\begin{equation}
    \overline{H}_N\ket{\Phi_0} = E_0 \ket{\Phi_0},
\end{equation}
where
\begin{equation}
 \overline{H}_N= e^{-T} H_N e^T
\end{equation}
is the similarity-transformed Hamiltonian obtained starting from
 $H_N$, which is normal-ordered Hamiltonian of Eq.~(\ref{H}) with respect to the reference $\ket{\Phi_0}$. In this work we use the normal-ordered Hamiltonian in the two-body approximation~\cite{hagen2007a,roth2012}.
Since the similarity-transformed Hamiltonian is non-Hermitian, the calculation of expectation values in CC theory requires the knowledge of both the left and right eigenstates. The right ground state is given by $\ket{0} = \ket{\Phi_0}$, while the left ground state is
\begin{equation}
    \bra{0} = \bra{\Phi_0}(1+\Lambda), \;\; \Lambda = \Lambda_1 + \Lambda_2 + \ldots,
\label{left_gs}
\end{equation}
where the operator $\Lambda$ is expanded as a sum of particle-hole de-excitation operators.

Our goal is to study also the dipole excitation of $^8$He and related sum rules. For this purpose, we  first
introduce the dipole response function
\begin{equation}
   R(\omega) =  \sum_{\mu} |\braket{\Psi_{\mu}|\Theta|\Psi_0}|^2 \delta(E_{\mu} - E_0 - \omega) \,,
\end{equation}
where $\ket{\Psi_{\mu}}$ are the excited states connected to the ground state by the dipole
operator  $\Theta$, and $\omega$ is the photon energy.
The dipole operator is given by
\begin{equation}
    \Theta = \sum_k^A (\mathbf{r}_k - \mathbf{R}_{CM})\left(\frac{1+\tau_k^3}{2}\right),
\end{equation}
where $\mathbf{r}_k$ and $\mathbf{R}_{CM}$ are the coordinates of the $k$-th nucleon and the center-of-mass, respectively, while $\tau_k^3$ is the third component of the isospin operator.

Due to the non-hermitian nature of the similarity transformed Hamiltonian, also for the excited states we have to distinguish between right and left eigenstates of the Hamiltonian. The latter are obtained using the CC equation-of-motion (EOM) method~\cite{stanton1993} and are defined as
\begin{eqnarray}
\label{eq:eomcc}
\nonumber
\overline{H}_N R_\mu \vert\Phi_0 \rangle = E_\mu R_\mu \vert \Phi_0
\rangle\, ,  \\
\langle \Phi_0 \vert L_\mu \overline{H}_N = E_\mu \langle \Phi_0\vert
L_\mu \, ,
\label{excited_states}
\end{eqnarray}
where the operators $R_{\mu}$ and $L_{\mu}$ are expressed in terms of a linear combination of particle-hole excitations as well.

Using Eq.\;(\ref{left_gs}) and Eq.\;(\ref{excited_states}), we can write the response function corresponding to the similarity-transformed Hamiltonian as
\begin{equation}
\begin{split}
   R(\omega) &= \sum_{\mu} \langle \Phi_0 \vert (1 + \Lambda ) \overline{\Theta}_N^\dagger R_\mu \vert \Phi_0 \rangle
\langle \Phi_0 \vert L_\mu \overline{\Theta}_N \vert \Phi_0 \rangle \\&\times \delta(E_{\mu} - E_0 - \omega).
\end{split}
\end{equation}
where
\begin{equation}
\label{thetabar}
\overline{\Theta}_N= e^{-T} \Theta_N e^T
\end{equation}
is the similarity transformed transition operator.
It is important to observe that the sum over $\mu$ in  $R(\omega)$ corresponds to both a sum over discrete excited states and an integral over continuum eigenstates of the Hamiltonian. In particular, the calculation of the latter represents a formidable task. Continuum state wavefunctions, in fact, contain information about all the possible fragmentation channels of the nucleus at a given energy.
To avoid the issue of explicitly computing the states in the continuum, we merged the CC method with the LIT technique~\cite{efros1994, efros2007}, originally used in few-body calculations. This led to the development of the so-called LIT-CC method, where the calculation of an integral transform of $R(\omega)$ with Lorentzian kernel
\begin{equation}
    L(\sigma,\Gamma) = \frac{\Gamma}{\pi} \int d\omega\; \frac{R(\omega)}{(\omega-\sigma)^2 + \Gamma^2}
\end{equation}
is directly related to the CC equation-of-motion method with a source term~\cite{bacca2013,bacca2014}. Once $L(\sigma,\Gamma)$ is calculated, a numerical inversion procedure allows one to recover $R(\omega)$ (see Ref.~\cite{efros2007} for details).

Starting from the LIT, one can easily obtain an estimate of the electromagnetic sum rules, i.e., the moments of the response function interpreted as a distribution function. Knowing all the moments is equivalent to  knowing the distribution itself. However, it is sometimes easier to compute just a few moments of a distribution rather than the full distribution, and yet
 obtain substantial insights into the dynamics of a quantum system.

 The moments (or sum rules) of the response function are defined as
\begin{equation}
    m_n = \int d\omega\;\omega^n R(\omega),
\end{equation}
where $n$ is an integer. Because in the limit $\Gamma\rightarrow 0$ the Lorentzian kernel becomes a delta function, we have that
\begin{equation}
    L(\sigma, \Gamma\rightarrow 0) = \int d\omega\;R(\omega)\delta(\omega -\sigma) = R(\sigma)\,,
\label{litsmallgamma}
\end{equation}
i.e., the moments can be computed from the LIT as
\begin{equation}
    m_n = \int d\sigma\; \sigma^n L(\sigma, \Gamma\rightarrow 0).
\label{mn_def}
\end{equation}
As shown in Ref.~\cite{miorelli2016}, this method is equivalent to obtaining first $R(\omega)$ and then integrating it, with the advantage that one does not have to perform an inversion, a procedure which can add to the
total numerical error budget.
 In this work, we will focus on the calculation of two dipole sum rules: the energy-weighted  $m_1$ and the inverse-energy weighted $m_{-1}$.


\section{Results}
\label{results}
\noindent
In this paper we  aim at performing a systematic study of the ground- and dipole excited-state properties of \ce{^8He}, supported by a reliable estimate of our theoretical uncertainties. Our computations are affected mainly by three sources of uncertainties: $(i)$  the model space truncation, $(ii)$ the many-body truncation, and $(iii)$ the dependence on the employed interaction model.

In order to address $(i)$, we need to take into account the fact that the expansion on the model space  is controlled by the maximum number of HO shells $N_{max}$ included in the calculation, for a given HO frequency $\hbar\Omega$.
For sufficiently large $N_{max}$ the results should be virtually independent of the choice of $\hbar\Omega$. In this work, we use $N_{max} = 14, 16$ and use the small residual $\hbar\Omega$-dependence (varying $\hbar\Omega$ in the range $12$ -- $16$ MeV) as a way to assess the uncertainty in the model space truncation.

In CC theory the particle-hole expansion of the cluster operator is truncated due to computational limitations. To address $(ii$) in this work we explore the effect of the CC truncation on both the ground and the excited states. In computing ground-state properties, the most frequently adopted approximation is CC with singles and doubles excitations, corresponding to $T$ = $T_1$ + $T_2$ and $\Lambda$ = $\Lambda_1$ + $\Lambda_2$. We will denote this truncation scheme with D. Next, we will also analyze results obtained by including leading order 3$p$-3$h$ excitations, which we denote here with T-1~\cite{watts1993}. In calculating excited-state observables, we will also make two choices for the approximation level of the EOM computation, either D or T-1~\cite{watts1995,jansen2016}.
Since the calculation of the dipole response functions and sum rules require us to perform a particle-hole expansion for the ground state ($T$ and $\Lambda$), and a corresponding one for the EOM computation of the excited states ($R_{\mu}$ and $L_{\mu}$), we indicate two expansion schemes. The resulting  CC truncation schemes for the dipole sum rule calculation are listed in Table \ref{tab:I}.
\begin{table}[h]
\caption{\label{tab:I}List of labels used to identify the CC truncation for the ground state (left of '/') and the excited states (right of '/'). Note that for Eq.~(\ref{thetabar}), we include only up to 2$p$-2$h$ excitations in the $T$, as  the effect of triples corrections is negligible~\cite{miorelli2018}.}
\begin{ruledtabular}
\begin{tabular}{lll}
 Ground state   & EOM calculation & Truncation scheme \\
 \hline
   D  & D & D/D \\
   T-1  & D & T-1/D \\
   T-1  & T-1 & T-1/T-1 \\
\end{tabular}
\end{ruledtabular}
\end{table}
We estimate the uncertainty $(ii)$ at the optimal HO  frequency $\hbar\Omega$ by computing the difference between the results obtained with two CC schemes.

Finally, to address $(iii)$, we compare the results obtained using three different chiral EFT interactions: \NNLOsat~\cite{ekstrom2015} and the $\Delta$-full interactions \NNLOgodlow\;and \NNLOgod~\cite{jiang2020}. These interactions are all given at next-to-next-to-leading order in the chiral expansion, and include three-body forces. It is worth noticing that the two $\Delta$-full interactions differ just by the value of the cutoff, which is $394$ MeV/c
and $450$ MeV/c for \NNLOgodlow\ and \NNLOgod\,, respectively. Employing these three different chiral interaction models enables us to appreciate the effect of including explicit delta isobars, varying the cutoff and the employed optimization protocol used for the LECs on the observables under analysis. A rigourous order-by-order treatment of $\chi$EFT uncertainty is left for future work.

In the results we will display in this Section, the uncertainties stemming from $(i)$ and $(ii)$ are added in quadrature, following Ref.~\cite{simonis2019}. Regarding $(iii)$, we work with three interactions and present them separately in Tables and Figures.



\subsection{Ground state properties}
\label{ground_state}

We start by presenting  CC results for the ground state energy $E_{gs}$ of \ce{^8He}. While  $E_{gs}$ was already obtained from coupled-cluster theory using the \NNLOsat\;interaction in Ref.~\cite{ekstrom2015}, in this work we further extend the analysis by showing the convergence with respect to the CC truncation, $N_{max}$ and $\hbar\Omega$, and by comparing the results from \NNLOsat\;with those from \NNLOgodlow\;and \NNLOgod.
\begin{figure}[hbt]
  \includegraphics[width=0.5\textwidth]{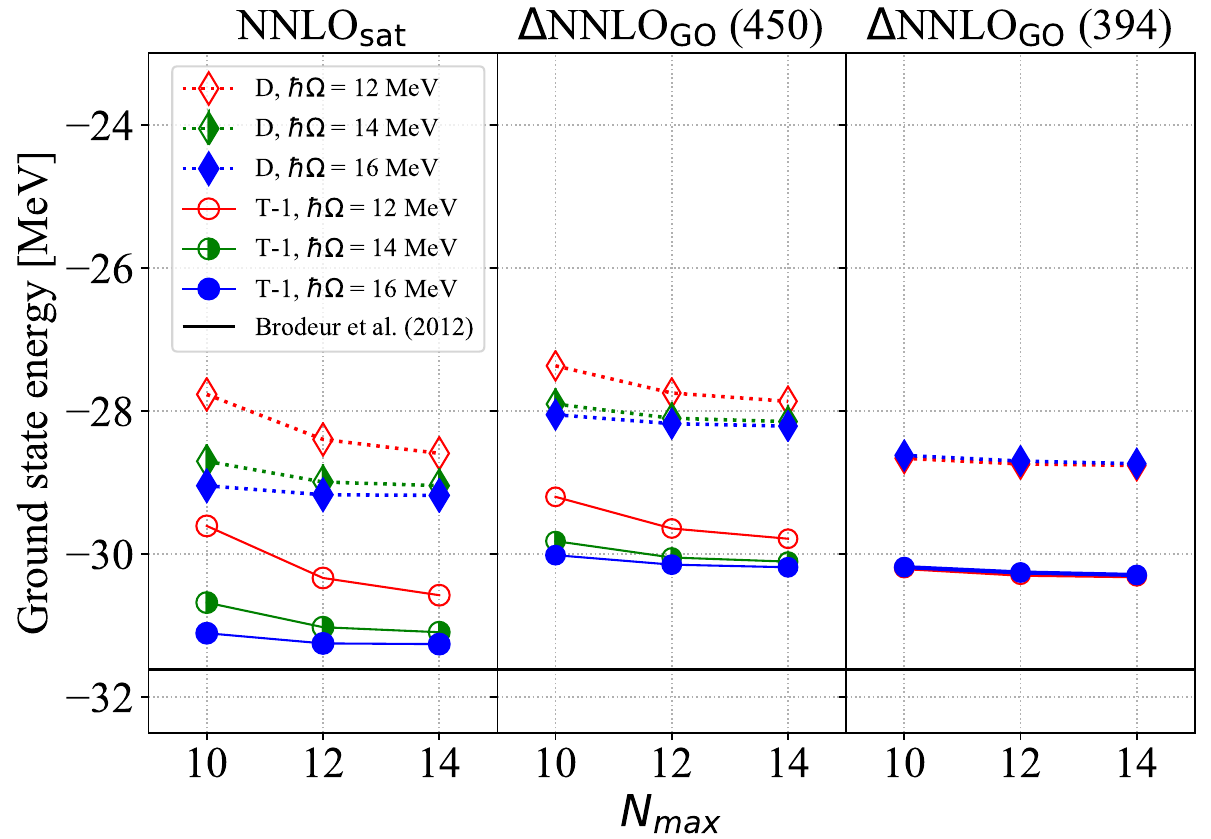}
  \caption{Ground state energy of \ce{^8He} in the D and T-1 scheme as a function of the model-size $N_{max}$ for the three Hamiltonians. The experimental value of the ground state energy is taken from Ref.~\cite{brodeur2012}.}
  \label{egs}
\end{figure}

The model space convergence for $E_{gs}$ with a comparison to experimental data is shown in Figure \ref{egs}, for the D and T-1 truncation schemes. At the optimal frequency $\hbar\Omega = 16$ MeV, the ground state energy values are converged with respect to $N_{max}$ for all three interactions. In the case of the $\Delta$-full interactions it is interesting to discuss the effect of the different values of the cutoff on the $\hbar\Omega$ dependence: while the \NNLOgodlow\;results for $\hbar\Omega = 12$ MeV and $\hbar\Omega = 16$ MeV are fully overlapped, for \NNLOgod\;the dependence on the HO frequency remains apparent.  This is expected, as the higher cutoff leads to a harder interaction.
The role of triples corrections is crucial in bringing theory closer to experiment. For all the interactions, 3$p$-3$h$ excitations represent between $8\%$ and $9\%$ of the CCSD correlation energy, moving our theoretical results in the direction of the experimental value. The final outcomes for the ground state energy are reported in Table \ref{tab:II}. We remark that the uncertainty, obtained summing $(i)$ and $(ii)$ in quadrature, is dominated by the CC truncation.  The \NNLOsat\;interaction gives the best agreement with respect to the experimental energy. We note that it has recently been shown that \ce{^8He} is soft towards being deformed in its ground-state~\cite{launey2020,holl2021}. The static correlations associated with deformation is not accurately captured in our spherical CC approach, and this might explain the slight underbinding we find for \NNLOgod\;and \NNLOgodlow.

\begin{table}
\caption{\label{tab:II} Theoretical predictions for the ground state energy of \ce{^8He} in MeV for the three different interactions in comparison to experiment.}
\begin{ruledtabular}
\begin{tabular}{ll}
 Interaction   & Ground state energy\\
 \hline
   \NNLOsat  & -31(1) \\
   \NNLOgod  & -30(1)  \\
   \NNLOgodlow  & -30.3(8) \\\\
   Experiment~\cite{brodeur2012} & -31.60972(11)\\
\end{tabular}
\end{ruledtabular}
\end{table}


\begin{figure}[hbt]
  \includegraphics[width=0.5\textwidth]{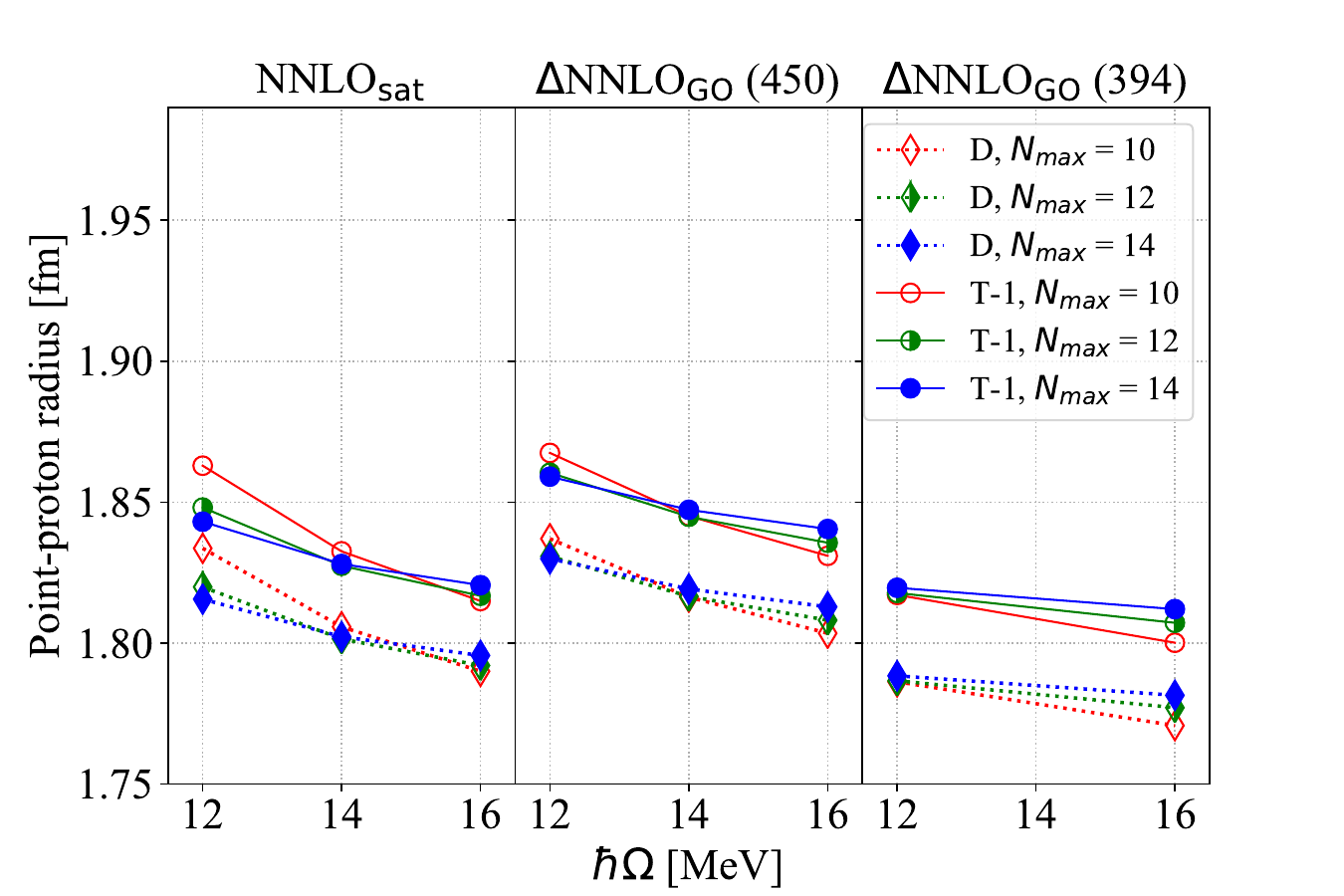}
  \caption{Point-proton radius of \ce{^8He} in the D and T-1 scheme  as function of the HO frequency $\hbar\Omega$ for the three Hamiltonians.}
  \label{rpp}
\end{figure}

Next we turn to our results for the charge radius of \ce{^8He}.
In order to analyze the nuclear charge radius $R_{ch}$, we first compute   the point-proton radius $R_{pp}$.
Coupled-cluster results of $R_{pp}$ as a function of $\hbar\Omega$ are shown in Fig.~\ref{rpp}, for the D and T-1 truncation schemes.
For each interaction, we identify the optimal frequency, leading to the best convergence as a function of the model space size, with the crossing point of the curves characterised by different $N_{max}$. In particular, for \NNLOsat\;and \NNLOgod\; (both with cutoff of $450$ MeV/c) the optimal frequency is $\hbar\Omega = 14$ MeV, while for \NNLOgodlow, the optimal frequency is $12$ MeV. As in the case of the ground state energy, we see that the cutoff affects the $\hbar\Omega$ dependence of our results, leading to a faster convergence when the cutoff is lower.
Moreover, we notice that the effect of triples corrections for $R_{pp}$ is smaller than in the case of $E_{gs}$. At the optimal frequency, in fact, the difference between the $R_{pp}$ values for D and T-1 amounts to around $1.5\%$ for all interactions.

Starting from $R_{pp}$, we then compute the charge radius
using
\begin{equation}
    \braket{R_{ch}^2} = \braket{R_{pp}^2} + R_p^2 + \frac{N}{Z} R_n^2 + \frac{3}{4M_p^2} + R_{so}^2 \,,
\end{equation}
where $R_p$ = $0.8414(19)$ \SI{}{fm}~\cite{codata2018} is the proton charge radius, $R_n^2$ = $-0.106^{+0.007}_{-0.005}$ \SI{}{fm^2}~\cite{filin2020} is the neutron charge radius, $3/(4M_p^2)$ = $0.033$ \SI{}{fm^2} is the Darwin-Foldy term and $R_{so}^2$ is the spin-orbit correction.
In Ref.~\cite{ong2010} it has been pointed out that $R_{so}^2$ could give a remarkable contribution to the charge radius of halo nuclei. Therefore, we have consistently calculated this correction in CC theory, improving in this respect Ref.~\cite{ekstrom2015}. Our results are reported in Table \ref{tab:III}, in comparison to previous theoretical estimates. Also in this case the uncertainties are obtained by summing in quadrature $(i)$
and $(ii)$.
\begin{table}
\caption{\label{tab:III} Theoretical predictions for the spin-orbit correction to the charge radius of \ce{^8He} for the three different interactions in comparison to previous theoretical results. \newline}
\begin{ruledtabular}
\begin{tabular}{ll}
 Interaction   & $R_{so}^2$ [\SI{}{fm^2}]\\
 \hline
   \NNLOsat  & -0.143(6) \\
   \NNLOgod  & -0.134(9)  \\
   \NNLOgodlow  & -0.141(6) \\\\
   Ref.~\cite{ong2010} & -0.17\\
   Ref.~\cite{papadimitriou2011} & -0.158\\
\end{tabular}
\end{ruledtabular}
\end{table}
The prediction of Ref.~\cite{ong2010} is based on a shell model calculation, while in Ref.~\cite{papadimitriou2011} the complex-energy configuration interaction method is employed. In this framework, the CC approach allows us to account for many-body correlations, leading to a significant improvement with respect to previous calculations of this quantity. In fact, in CC the magnitude of $R_{so}^2$ is  reduced of about $10\%$ with respect to Ref.~\cite{papadimitriou2011}, and of approximately $20\%$ with respect to the shell model estimate.

\begin{figure}[hbt]
   \includegraphics[width=0.5\textwidth]{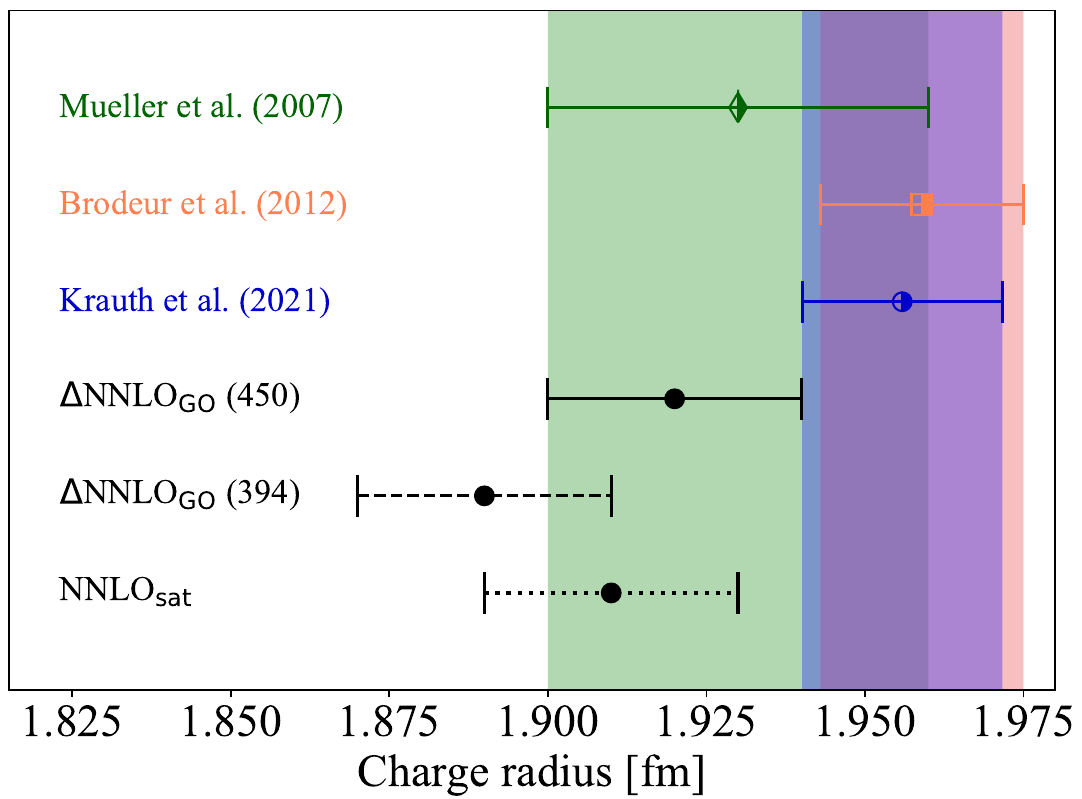}
   \caption{Comparison between the coupled-cluster theoretical values for the charge radius of \ce{^8He} using the three different Hamiltonians and the experimental results of Mueller et al.~\cite{mueller2007}, Brodeur et al.~\cite{brodeur2012} and Krauth et al.~\cite{krauth2021}.}
   \label{rch}
\end{figure}

The final results for the charge radius of \ce{^8He} are illustrated in Figure \ref{rch} in comparison to three  experimental determinations. The charge radius of \ce{^8He} can be experimentally obtained from a measurement of the isotope shift, namely the frequency difference $\delta \nu_{\rm A,A'}$  between \ce{^8He} and the reference isotope \ce{^4He}, in the same atomic transition. The frequency shift is related to the difference $\delta \braket{R^2_{ch}}_{\rm A,A'}$ in the charge radius  between \ce{^8He} and \ce{^4He}  by
\begin{equation}
\delta \nu_{\rm A,A'} = \delta^{Mass}_{\rm A,A'} +
K_{FS} \delta \braket{R^2_{ch}}_{\rm A,A'}\,,
\end{equation}
where the mass shift $\delta^{Mass}_{\rm A,A'}$ and the field shift constant $K_{FS}$ are obtained from
precise atomic theory calculations.
The first determination of the charge radius of \ce{^8He}
using this method stems from Ref.~\cite{mueller2007}, where the radius of \ce{^4He} measured from electron scattering was used as a reference. Later, Ref.~\cite{brodeur2012} provided an improved estimate of the mass and field shift parameters, based on precise nuclear mass measurements. More recently,  Ref.~\cite{krauth2021} achieved the first determination of the \ce{^4He} charge radius from muonic atoms,
which, improving the  reference, slightly modified  the $R_{ch}$ for  \ce{^8He}.

In general, we find that our theoretical results are in good accordance with the experimental determinations, as seen in Fig.~\ref{rch}. In particular, the \NNLOgod\;interaction leads to the largest charge radius, equal to $1.92(2)$ fm, which agrees best with \cite{mueller2007}. The larger radius is due to the interplay between the higher value for $R_{pp}$ and the smaller value of $R_{so}^2$ obtained with this interaction in comparison to the other two. Moreover, the distance between the upper end of \NNLOgod\;error bar and the lower end of the one of the most recent experimental determination~\cite{krauth2021} amounts to $10^{-4}$ fm. Comparing this to the scale of the values involved, we can still claim a good agreement between these two results.

\subsection{Discretized dipole response function  and energy-weighted sum rule}
\label{dipole}

\begin{figure}[hbt]
  \includegraphics[width=0.5\textwidth]{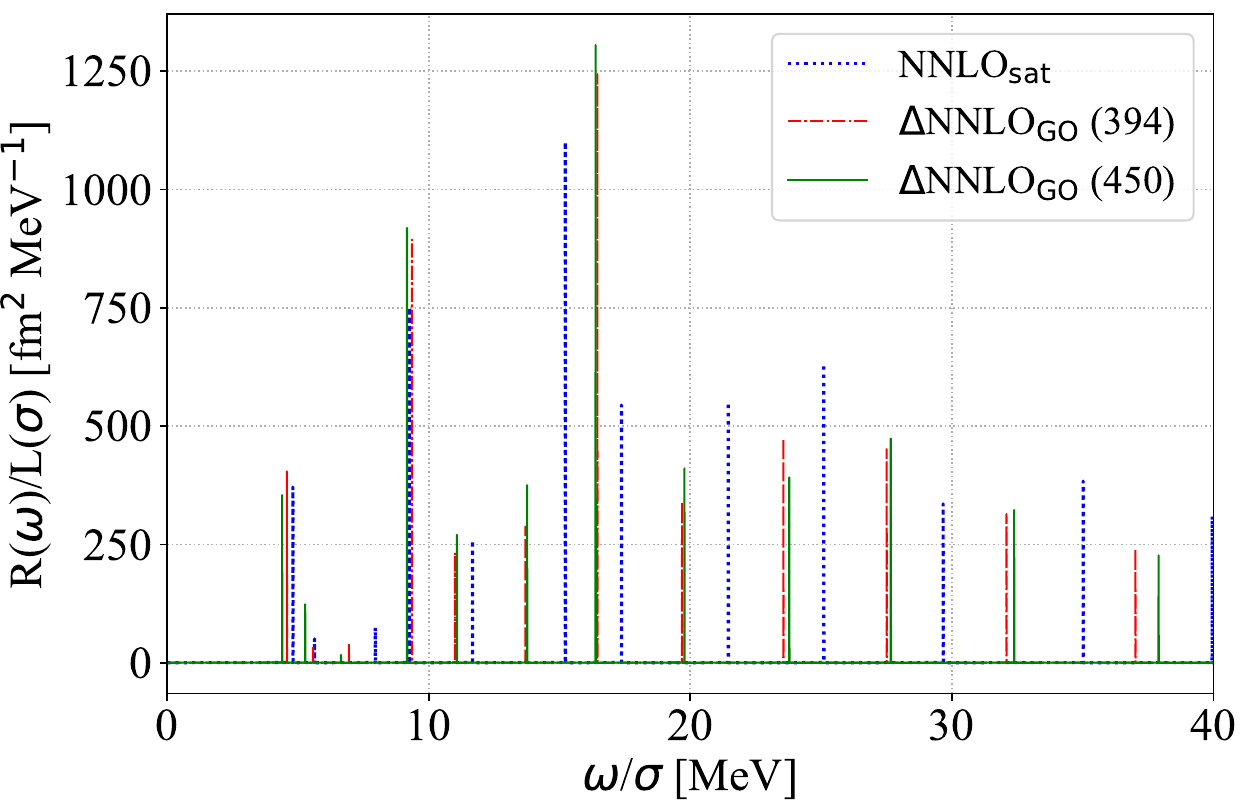}
  \caption{Discretized response function of \ce{^8He} with $\Gamma = 10^{-4}$ MeV in the T-1/T-1 scheme for the different chiral EFT interactions.}
  \label{lit}
\end{figure}

We now address the dipole excited states in ${^8}$He by first looking at the discretized response function.
 In our framework, this quantity can be simply calculated taking the limit of the LIT for $\Gamma\rightarrow 0$, as shown in Eq.~(\ref{litsmallgamma}).
In Figure \ref{lit}, we show the LIT with $\Gamma = 10^{-4}$ MeV for the three different interactions.
For all the potentials, the discretized response function presents low-energy peaks emerging at around $5$ MeV. On the one hand, our results are consistent with the analysis of the $^3$H($^6$He,$p$)$^8$He transfer reaction in Ref.~\cite{golovkov2009} where a low-lying dipole strength around $3$~MeV was indicated. On the other hand, the recent inelastic proton scattering experiment on $^8$He~\cite{holl2021} did not observe any low-lying dipole resonance below $5$~MeV.
The larger number of states that we observe at about 20 MeV for all three interactions correspond to the giant dipole resonance.

\begin{figure}[hbt]
  \includegraphics[width=0.5\textwidth]{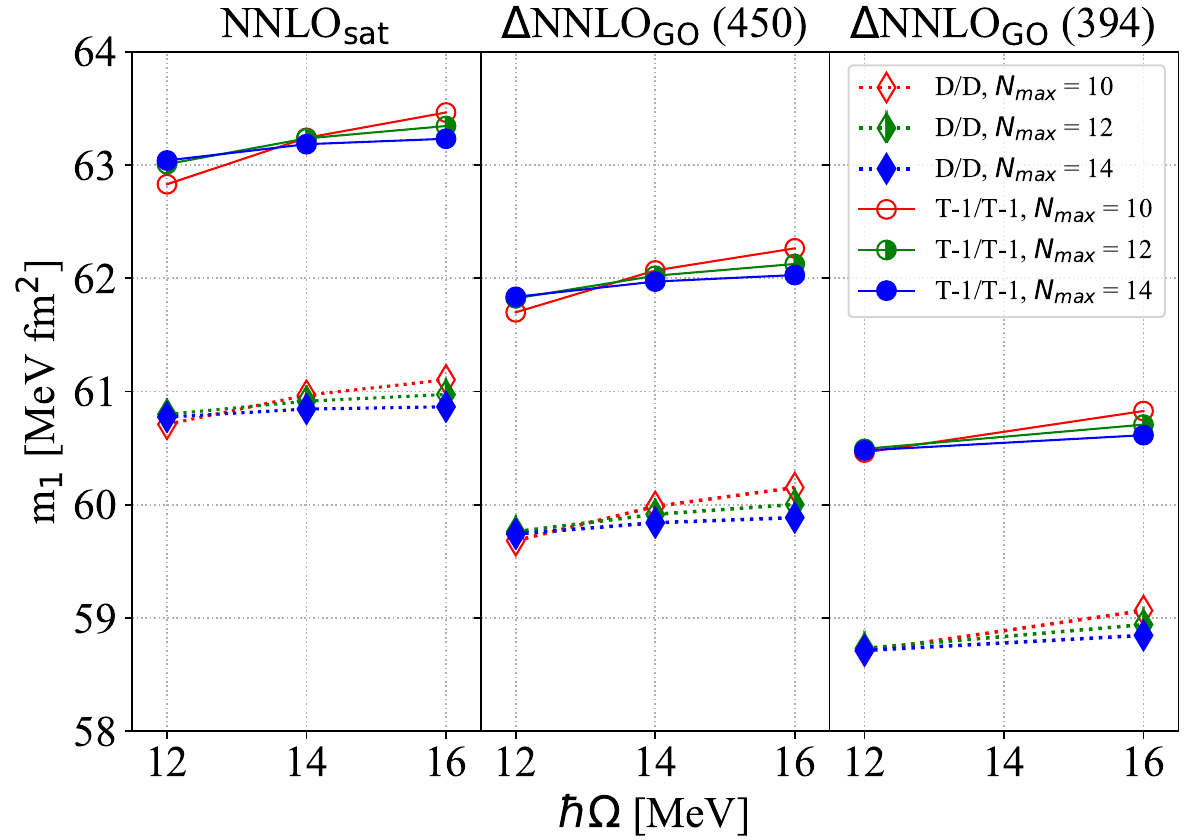}
  \caption{Energy-weighted dipole sum rule for \ce{^8He} as a function of  $\hbar\Omega$ for the three Hamiltonians in the D/D and T-1/T-1 truncation scheme. }
  \label{m1}
\end{figure}

Starting from the discretized response function and using
Eq.~(\ref{mn_def}), we can study the energy-weighted dipole sum rule $m_1$.  Figure \ref{m1} illustrates its behavior as a function of $\hbar\Omega$ for various $N_{max}$ in the D/D and T-1/T-1 truncation scheme.
At a first glance, we immediately notice that the energy-weighted sum rule converges quite quickly.  In the case of $m_1$, the largest contribution to the integral of Eq.~(\ref{mn_def}) is represented by the high energy dipole-excited  states, which are well converged.
It is interesting to look at the effect of varying the interaction cutoff on the $\hbar\Omega$ convergence pattern: while for \NNLOsat\;and \NNLOgod\; we achieve a negligible dependence on $N_{max}$ at $\hbar\Omega = 14$ MeV, in the case of \NNLOgodlow\;the optimal frequency corresponds to $\hbar\Omega = 12$ MeV.

The contribution of triples, going from the D/D to the T-1/T-1 truncation scheme is substantial and amounts to $3.5\%$ for \NNLOsat\;and \NNLOgod, and to $3\%$ for \NNLOgodlow.
However, we have noticed that the differences between the T-1/D (not shown in  Figure \ref{m1}) and T-1/T-1 results are of the order of $0.01\%$,  hence negligible.
This fact can be understood if we consider an alternative expression for the energy-weighted sum rule. Let us assume we can write the similarity transformed Hamiltonian $\overline{H}_N$ in a tridiagonal form using the Lanczos algorithm~\cite{lanczos1950}. Choosing the vectors
\begin{equation}
    \bra{w_0} = \frac{\bra{0}\overline{\Theta}^{\dagger}_N}{\sqrt{\braket{0|\overline{\Theta}^{\dagger}_N\overline{\Theta}_N|0}}}, \quad
    \ket{v_0} = \frac{\overline{\Theta}_N\ket{0}}{\sqrt{\braket{0|\overline{\Theta}^{\dagger}_N\overline{\Theta}_N|0}}}
\end{equation}
as the left and right pivots, respectively, the energy-weighted sum rule can be calculated as the product between the non-energy-weighted sum rule $m_0 = \braket{0|\overline{\Theta}^{\dagger}_N\overline{\Theta}_N|0}$ and the first Lanczos coefficient $a_0 = \braket{w_0|\overline{H}_N|v_0}$. Because $m_0$ is a pure ground state expectation value,  it is the same in both the T-1/D and T-1/T-1 schemes. Therefore, triples contributions from the EOM calculation enter in $m_1$ only via $a_0$, explaining the small difference between the T-1/D and T-1/T-1 scheme.

Our result for the energy-weighted sum rule can be compared to the Thomas-Reiche-Kuhn (TRK) sum rule, which is often discussed when studying photoabsorption cross sections $\sigma_{\gamma}(\omega)$~\cite{orlandini1991}. The TRK sum rule is defined as
\begin{equation}
    \int_{\omega_{th}}^{\infty} d\omega\; \sigma_{\gamma}(\omega) = 5.974 \frac{NZ}{A}\; \SI{}{MeV\;fm^2}\;
    (1+\kappa)\,,
    \label{trk}
\end{equation}
where $\omega_{th}$ is the threshold energy and $\kappa$ is the so-called enhancement factor~\cite{eisenberg1976}. The latter arises from the presence of exchange terms in the nuclear force, which do not commute with the dipole operator~\cite{orlandini1991}. Considering that $\sigma_{\gamma}(\omega) = 4\pi^2\alpha \omega R(\omega)$, we can connect $m_1$ to the left-hand side of Eq. (\ref{trk}) according to
\begin{equation}
     \int_{\omega_{th}}^{\infty} d\omega\; \sigma_{\gamma}(\omega) = 4\pi^2 \alpha m_1.
\label{sigma_m1}
\end{equation}
Combining Eq.~(\ref{trk}) and Eq.~(\ref{sigma_m1}), we are then able to evaluate the enhancement factor for the employed interactions. Our results for $\kappa$, accompanied by our final estimates for $m_1$, are shown in Table \ref{tab:V}.  Also in this case uncertainties are obtained summing in quadrature the contributions of $(i)$ and $(ii)$.
To be more conservative in the uncertainty estimate, the CC truncation uncertainty $(ii)$ on $m_1$ has been calculated comparing the T-1/T-1 results directly with the D/D ones, instead of considering the variations between the two best CC schemes available (T-1/T-1 and T-1/D).

\begin{table}
\caption{\label{tab:V} Theoretical predictions for the energy-weighted dipole sum rule of \ce{^8He}  and the enhancement factor for the three different chiral EFT interactions under consideration. \newline}
\begin{ruledtabular}
\begin{tabular}{lll}
\label{kappa}
 Interaction   & $m_1$ [\SI{}{MeV\;fm^2}]& $\kappa$\\
 \hline
       \NNLOsat  & 63(1) & 1.02(3) \\
       \NNLOgod & 62(1) &  1.00(3) \\
       \NNLOgodlow & 60.5(9) & 0.94(3) \\
\end{tabular}
\end{ruledtabular}
\end{table}

We observe that \NNLOsat\;produces the largest value of $m_1$ and consequently of $\kappa$. The three interactions are all consistent in predicting a quite large enhancement factor ranging between $0.9$ and $1$. This is not surprising as there are components of non-locality in these interactions, both at the nucleon-nucleon level and at the three-nucleon force level. Large enhancement factors of about 0.6 had already been observed in Ref.~\cite{bacca2014}, where only two-body forces were used.

Further insight on the E1 strength function can be obtained looking at the cluster sum rule~\cite{alhassid1982, hencken2004}. This quantity is associated with the excitation of a "molecular" dipole degree of freedom, related to the relative motion of two clusters inside the nucleus. Indicating with $(A_i,\; Z_i)$, $i = 1,\;2$ the mass and proton number of  the two clusters, and with $(A,\;Z)$ those of the nucleus, we can connect the TRK sum rule to the cluster sum rule  as
\begin{eqnarray}
    \label{cluster}
    \int_{\omega_{th}}^{\infty} d\omega\; [\sigma_{\gamma}(\omega) - \sigma^{cl_1}_{\gamma}(\omega) - \sigma^{cl_2}_{\gamma}(\omega)] = \\5.974
    \nonumber
    \frac{(Z_1 A_2 – Z_2 A_1)^2}{AA_1 A_2} \SI{}{MeV\;fm^2}\; (1+\kappa).
\end{eqnarray}

The evaluation of the cluster sum rule becomes particularly interesting for a halo nucleus such as \ce{^8He}. The latter, in fact, can be modelled as a \ce{^4He} core, representing the first cluster $(cl_1)$, surrounded by four neutrons, constituting the second cluster ($cl_2$). By choosing $Z_1=2,~ Z_2=0$
and $A=8,~A_1=4,~A_2=4$ one can compute the r.h.s.~of Eq.~(\ref{cluster}) and compare it to the l.h.s.,
which is obtained by integrating the photoabsoroption cross section of $^{8}$He and subtracting that of $^{4}$He given that
$\sigma_{\gamma}^{cl_2}=0$.

For the \NNLOsat\;interaction, computing  the energy-weighted dipole sum rule of \ce{^4He} in the CC framework, we obtain $41.9(3)$ \SI{}{MeV\;fm^2}, which using Eq.~(\ref{trk}), yields an enhancement factor of $1.02(2)$, compatible with the results in Table~\ref{kappa}.
The cluster sum rule for the \NNLOsat\;interaction becomes then
\begin{equation}
    \int_{\omega_{th}}^{\infty} d\omega\; [\sigma_{\gamma}^{^8{\rm He}}(\omega) - \sigma_{\gamma}^{^4{\rm He}}(\omega) ] = 6.0(3) \SI{}{MeV\;fm^2}.
\end{equation}
By looking at the ratio of the cluster sum rule with respect to the TRK sum rule, we are able to quantify how much of the dipole strength of \ce{^8He} is given by the relative motion between core and halo. The latter turns out to be approximately $30\%$, which means that the core-halo relative motion appears to determine around $1/3$ of the total E1 strength for \ce{^8He}.


\subsection{Electric dipole polarizability}

Finally we turn to our calculations of the electric dipole polarizability $\alpha_D$ for \ce{^8He}.
The latter is related to the inverse-energy weighted sum rule  by
\begin{equation}
    \alpha_D = 2\alpha \int d\omega\; \frac{R(\omega)}{\omega} = 2\alpha m_{-1}\,,
\label{alphaD_m-1}
\end{equation}
where $m_{-1}$ is calculated using Eq.~(\ref{mn_def}).
This implies that $\alpha_D$ is mainly determined by the low-energy part of the discretized spectrum of Fig.~\ref{lit}, in particular by the first states at about 5 MeV.

\begin{figure}[hbt]
  \includegraphics[width=0.5\textwidth]{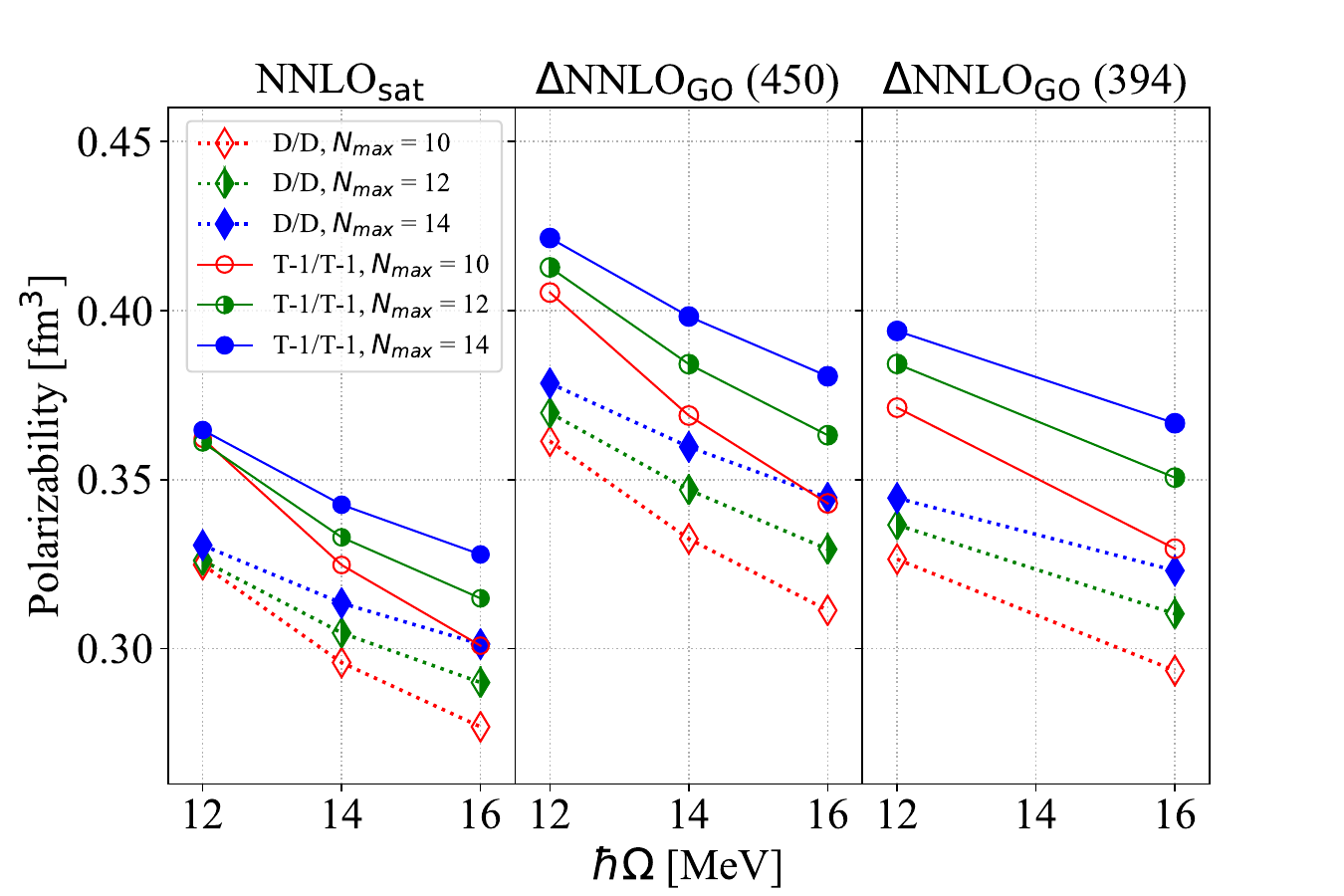}
  \caption{Dipole polarizability of \ce{^8He} as a function of the harmonic oscillator frequency $\hbar\Omega$ for the three Hamiltonians, in the D/D and T-1/T-1 truncation schemes.}
  \label{pol}
\end{figure}
\noindent

In Figure \ref{pol}, we show the convergence pattern of  $\alpha_D$ with respect to $\hbar\Omega$, in the D/D and T-1/T-1 truncation schemes.
For all three interactions we observe a quite pronounced dependence of the results on the CC truncation and model space parameters. Variations with respect to $N_{max}$ tend to reduce in correspondence to small values of the HO frequency.
This slow convergence of the polarizability is probably related to the slow convergence of low-lying dipole states.

Triples corrections give a significant contribution to $\alpha_D$ both in the ground and excited state part of the CC calculation. For $\hbar\Omega = 12$ MeV, the inclusion of 3$p$-3$h$ excitations just in the ground state (T-1/D scheme, not shown in  Figure~\ref{pol}) leads to an increase of $\alpha_D$ between $3\%$ and $5\%$ with respect to the D/D scheme results. If we add triples also in the EOM calculation (T-1/T-1 shown in Figure~\ref{pol}), we achieve an overall $10\%$ enhancement of the polarizability compared to the full doubles computation.
This is different with respect to what observed for $m_1$.
The reason is that, as shown in Ref.~\cite{miorelli2016}, $\alpha_D$ can be rewritten as a continued fraction involving the whole set of Lanczos coefficients available, while $m_1$  just depends on the first one of them, $a_0$.

Our final results for the dipole polarizability are reported in Table \ref{tab:IV}. Uncertainties are obtained by summing in quadrature $(i)$ and $(ii)$, where the CC truncation error is computed as the difference between the results of the T-1/T-1 and D/D schemes.
Our theoretical uncertainty varies between $7\%$ and $10\%$ of the central value for the different interactions. It is worth noticing that with respect to the other observables previously discussed, we get a more conservative estimate for the uncertainty, which reflects the slow convergence for $\alpha_D$. This might be due to the loosely bound halo neutrons in \ce{^8He}, which determine a more extended wavefunction, and as a consequence a slower convergence.
Finally, the fact that the \NNLOgod\; yields the largest prediction for $\alpha_D$ as seen in Table \ref{tab:IV} is related to the dipole strength showing a state at slightly lower energies with respect to  the other interactions, see Fig.~\ref{lit}.

 \begin{table}
\caption{\label{tab:IV} Theoretical predictions for the dipole polarizability of \ce{^8He} for the three Hamiltonians. \newline}
\begin{ruledtabular}
\begin{tabular}{ll}
 Interaction   & $\alpha_D$ [\SI{}{fm^3}]\\
 \hline
   \NNLOsat  & 0.37(3) \\
   \NNLOgod  & 0.42(3)  \\
   \NNLOgodlow  & 0.39(2) \\
\end{tabular}
\end{ruledtabular}
\end{table}

Interestingly, our calculations show that the dipole polarizability of \ce{^8He} is more than five times larger than that of \ce{^4He}. Combining the photoabsorption cross section data of Refs.~\cite{arkatov1974,arkatov1980, pachucki2007} the latter amounts to $0.074(9)$\;\SI{}{fm^3}, which is compatible with the results of Ref.~\cite{miorelli2016}.
A larger polarizability is expected in halo nuclei due to soft dipole mode excitations, such as those shown in Figure~\ref{lit} at about 5 MeV, which are not seen in  \ce{^4He}~\cite{bacca2014,miorelli2016}.

\section{Conclusions}
\label{conclusions}
\noindent
In this paper we carried out a systematic investigation of ground and dipole excited states of the \ce{^8He} halo nucleus using CC theory and different chiral nucleon-nucleon and three-nucleon forces. We obtain results that are  consistent with experiment for the ground-state energy, albeit with a slight under-binding, and the nuclear charge radius.
Looking at the three implemented Hamiltonians separately, we see that the $\Delta$-full interaction \NNLOgod\; delivers the best agreement with the most recent experimental update of the charge radius of \ce{^8He}~\cite{krauth2021}.

We also presented the first theoretical predictions for the  energy-weighted  and inverse energy-weighted sum rules of \ce{^8He}.
From the former we compare the TRK and cluster sum rules to conclude that about 1/3 of the dipole strength is due to the excitation of the
molecular dipole degrees of freedom. For the latter, we provide a prediction that could be tested in future experiments. An experimental determination of the dipole strength of \ce{^8He}, performed with Coulomb excitation at RIKEN in Japan, is currently under analysis~\cite{aumann}.

To address the recent theoretical and experimental indications of deformation in the ground-state of $^8$He~\cite{launey2020,holl2021} we plan to extend the CC calculations reported in this work by starting from an axially symmetric reference state following Ref.~\cite{novario2020}. We also plan to address EFT truncation errors by performing an order-by-order study of the $\chi$EFT uncertainty, supported by Bayesian statistical tools~\cite{Furnstahl_2015}.

\begin{acknowledgments}
\noindent
This work was supported by the Deutsche Forschungsgemeinschaft  (DFG, German Research Foundation) through Project-ID 279384907 - SFB 1245 and through the Cluster of Excellence ``Precision Physics, Fundamental Interactions, and Structure of Matter" (PRISMA$^+$ EXC 2118/1, Project ID 39083149), by the Office of Nuclear Physics, U.S. Department of Energy, under grant DE-SC0018223 (NUCLEI SciDAC-4 collaboration), and contract No. DE-AC05-00OR22725 with UT-Battelle, LLC (Oak Ridge National Laboratory). Computer time was provided by the Innovative and Novel Computational Impact on Theory and Experiment (INCITE) program and by the supercomputer Mogon at Johannes Gutenberg Universit\"at Mainz. This research used resources of the Oak Ridge Leadership Computing Facility located at ORNL, which is supported by the Office of Science of the Department of Energy under Contract No. DE-AC05-00OR22725.
\end{acknowledgments}

\FloatBarrier 
\bibliography{refs,master} 

\end{document}